\documentclass[preprint,authoryear,review,12pt]{elsarticle}

\usepackage{amssymb}
\usepackage{amsmath}
\usepackage{graphicx}

\journal{Finance Research Letters}

\begin{document}

\begin{frontmatter}



\title{Can Analysts Predict Rallies Better Than Crashes?}


\author{Ivan Medovikov}

\address{Department of Economics, Brock University \\ 429 Plaza Building, 500 Glenridge Avenue, St. Catharines, Ontario, L2S 3A1, Canada.}

\begin{abstract}
We use the copula approach to study the structure of dependence between sell-side analysts' consensus recommendations and subsequent security returns, with a focus on asymmetric tail dependence. We match monthly vintages of I/B/E/S recommendations for the period January to December 2011 with excess security returns during six months following recommendation issue. Using a symmetrized Joe-Clayton Copula (SJC) model we find evidence to suggest that analysts can identify stocks that will substantially outperform, but not underperform relative to the market, and that their predictive ability is conditional on recommendation changes.
\end{abstract}


\end{frontmatter}


\section{Introduction}
\label{section_introduction}

Investment value of analyst recommendations has been the subject of considerable research during the past twenty years. This comes as no surprise, given the substantial resources that investment banks, brokerage houses and their clients spend on security analysts, and the attention that recommendations attract from the media and the investing public. It is now generally established that analysts possess stock-picking abilities, meaning trading strategies based on recommendations that yield positive returns in excess of the market are possible (see \cite{Stickel1995}, \cite{Womack1996}, \cite{Barber2001} and \cite{Jegadeesh2006}, among others), but the circumstances under which this is the case are not always clear. Profitability of a recommendation appears to depend on many factors, including whether it represents a revision or reiteration of an earlier opinion (\cite{Jegadeesh2004}, \cite{Barber2010}), on timely access to analyst reports (\cite{Green2006}), industry (\cite{Boni2006}), portfolio turnover (\cite{Barber2001}) and proximity of earnings announcements (\cite{Ivkovic2004}). 

While determinants of recommendation profitability received substantial attention, interestingly, no studies appear to focus on return characteristics associated with greater predictability. For example, are analysts better at calling out extreme price fluctuations than normal movements? If so, are there asymmetries in this relationship: can analysts better predict significant rallies or crashes? The aim of this paper is to fill this gap. 

The absence of literature on nonlinear and extreme dependence between ratings and returns may stem from the lack of suitable multivariate distribution functions that can accommodate the very different marginal behaviour of recommendations and returns, with return distributions  known to be symmetric and leptokurtic and  recommendation distributions - skewed and often bimodal. To this end, this paper adopts the copula approach and uses a highly-flexible semiparametric model to measure dependence between the level of recommendations and subsequent security returns. To the best of our knowledge, this appears to be the first application of copulas to the analysis of recommendations in the literature. 

Our particular focus is on tail dependence, or dependence among extremes. We match monthly vintages of I/B/E/S consensus recommendations issued during 2011 with corresponding excess security returns six months following recommendation issue, and using a symmetrized Joe-Clayton copula model find strong evidence of dependence in the upper, but not lower tail of the joint distribution. In other words, we find that stocks with most favorable recommendations tend to substantially outperform the market, while stocks with most unfavorable recommendations show no tendency to underperform, suggesting that analysts' predictive abilities are asymmetric between the extremes of the ratings distribution and are skewed toward picking substantially undervalued rather than overvalued stocks. We also find that this relationship only holds for stocks that experienced a recent deterioration of consensus opinion, suggesting that profit opportunities identified here may be driven by investors' over-reaction to declines in analysts' outlook for top-rated securities.

The paper is organized as follows. Section \ref{sec::methodology} reviews some basic concepts behind copulas, and introduces the copula model and estimation technique used in this paper. The matching of I/B/E/S data with excess returns as well as the filtering of the data is discussed in Section \ref{sec::data}. Estimation results are presented in Section \ref{sec::results}, followed by a brief discussion in Section \ref{sec::discussion}.

\section{The methodology}
\label{sec::methodology}

\subsection{Copula functions}

The copula approach is central to this paper, and we begin by reviewing some of the basic concepts behind the theory of copulas. Consider a pair of random variables $X$ and $Y$, and let $F(x)$ and $G(y)$ represent their marginal distribution functions (d.f.-s), and $H(x,y)$ be the joint d.f. Following a result by \cite{sklar}, the joint d.f. $H$ can be written as 
\begin{equation}
H(x,y) = C[F(x),G(y)],\hspace{4mm}(x,y) \in \mathbb{R}^2,
\end{equation} where the function $C:[0,1]^2 \rightarrow [0,1]$ is the so-called copula of $X$ and $Y$. Copulas have become central to the analysis of dependence as they provide a complete, and in the case of continuous random variables, a unique description of the relationship between $X$ and $Y$. Letting $u = F(x)$ and $v = G(y)$, it becomes clear that the copula is simply the joint d.f. of $(u,v)$ which we can write as $C(u,v) = H(F^{-1}(u),G^{-1}(v)),\hspace{4mm}(u,v) \in [0,1]^2$. Note that for any $F$ and $G$, $u$ and $v$ are uniform on $[0,1]$, meaning that the model of dependence encoded in $C$ is free from the specification of the marginals. A model of $H$ can therefore be constructed by specifying the marginals and the dependence structure separately. This feature makes copulas particularly well-suited for the analysis of dependence between recommendations and security returns, since it allows for easy combination of marginals that are very different. For an introduction to copulas see \cite{Joe1997}, \cite{Nelsen2006} and \cite{Cherubini2004} and \cite{Patton2009} for applications of copulas in finance.

The focus in this paper is on tail dependence, or dependence among extremes, which refers to the tendency of extremely large (or small) values of one variable to be associated with extremely large (or small) values of another. Such dependence is usually studied through upper- and lower-tail dependence coefficients denoted $\lambda_u$ and $\lambda_l$ respectively, and defined as:
\begin{eqnarray}
\lambda_u &=& \lim_{u\rightarrow1} Pr[F(x)\geq u | G(y) \geq v] = \lim_{u \rightarrow 1} \frac{1 - 2u + C(u,u)}{1-u} \\
\lambda_l &=& \lim_{u\rightarrow0} Pr[F(x)\leq u | G(y) \leq v] = \lim_{u \rightarrow 1} \frac{C(u,u)}{u}.
\end{eqnarray} Larger values of $\lambda_u$ ($\lambda_l$) indicate greater tendency of the data to cluster in the upper-right (lower-left) tail of the joint distribution, in which case the variables are said to be upper (lower) tail-dependent. The case of $\lambda_u=0$ and $\lambda_l = 0$ corresponds to the absence of dependence in the tails. Our aim here is to obtain estimates of coefficients of tail-dependence between the level of analyst consensus recommendations and subsequent security excess returns and to test for their significance and possible difference. 

\subsection{Symmetrized Joe-Clayton copula}

Asymmetric tail dependence between financial series has attracted some recent attention in the literature. \cite{Patton2006} and \cite{Michelis2010} use copulas to test for asymmetric exchange rate dependence and \cite{Ning2009} adopt a copula model to probe for asymmetries in tail dependence between equity index returns and trading volume. Here, we adopt a modeling approach similar to that of \cite{Ning2009} and \cite{Michelis2010}. To capture the asymmetries in the return-recommendation relationship a sufficiently-flexible model for $C$ is required. Interestingly, most common parametric copula models either allow for no tail dependence at all (e.g. Gaussian and Frank copulas), or restrict dependence to only one tail (e.g. Gumbel and Clayton copulas), or force dependence to be the same in all tails (e.g. t-copula). We employ the unconditional symmetrized Joe-Clayton (SJC) copula model of \cite{Patton2006}, which is defined as
\begin{align}
C_{SJC} (u,v | \lambda_u, \lambda_l) &= 0.5 \times (C_{JC}(u,v | \lambda_u, \lambda_l) + \\
& C_{JC}(1-u,1-v | \lambda_u, \lambda_l) +u + v -1),
\end{align} where $C_{JC}(u,v | \lambda_u, \lambda_l)$ is the Joe-Clayton (or BB7) copula given by
\begin{equation}
\label{eq::jccopula}
C_{JC}(u,v | \lambda_u, \lambda_l) = 1 - (1-\{[1-(1-u)^k]^{-r} + [1-(1-v)^k]^{-r} -1\}^{-1/r})^{1/k},
\end{equation} with $k = 1/\log_{2}(2-\lambda_u)$, $r = -1/\log_2(\lambda_l)$, and tail-dependence parameters $\lambda_u$ and $\lambda_l \in (0,1)$ defined as before. The SJC copula allows for both upper and lower-tail dependence, with the values of $\lambda_u$ and $\lambda_l$ determined independently from one another.

\subsection{Estimation}

The SJC \textit{copula density} is given by 
\begin{align}
c_{SJC}(u,v &| \lambda_u, \lambda_l) = \frac{\partial^2 C_{SJC} (u,v | \lambda_u, \lambda_l)}{\partial u \partial v} \\
& = 0.5 \times \left [ \frac{\partial^2 C_{JC}(u,v | \lambda_u, \lambda_l) }{\partial u \partial v} + \frac{\partial^2 C_{JC}(1-u,1-v | \lambda_u, \lambda_l) }{\partial (1-u) \partial (1-v)} \right ].
\end{align} It is relatively straightforward to obtain the partial $\partial^2 C_{JC}(u,v | \lambda_u, \lambda_l) / \partial u \partial v$ from (\ref{eq::jccopula}), and interested readers are referred to Section 4.2.1 of \cite{Michelis2010} for full expression. Estimates $\hat{\lambda}_u$ and $\hat{\lambda}_l$ can be obtained by maximizing the corresponding \textit{copula log-likelihood} function $l_c = \log(c_{SJC}(u,v | \lambda_u, \lambda_l))$ with respect to $\lambda_u$ and  $\lambda_l$. To avoid distributional assumptions, the marginals $F$ and $G$ can be replaced by their empirical counter-parts based on $n$ independent copies $(x_1,y_x),...,(x_n,y_n)$ that are given by:
\begin{equation}
\hat{F}(x) = \frac{1}{n} \sum_{i=1}^n \mathbb{I}(x_i \leq x) \text{ and } \hat{G}(y) = \frac{1}{n} \sum_{i=1}^n \mathbb{I}(y_i \leq y),
\end{equation} where $\mathbb{I}()$ is the indicator function. Uniform convergence of $\hat{F}(x)$ and $\hat{G}(y)$ to $F$ and $G$ follows from Glivento-Cantelli theorem. This immediately yields estimates $\hat{u}_i = \hat{F}(x_i)$ and $\hat{v}_i = \hat{G}(y_i)$, for $i=1..n$, and leads to the \textit{sample copula log-likelihood} function $L_c = \sum_{i=1}^n l_c(\hat{u}_i,\hat{v}_i | \lambda_u, \lambda_l)$ that is to be maximized. In practice, the empirical d.f-s are also rescaled by a factor $n/(n+1)$ to ensure that for any finite $x$ and $y$, $\hat{F}(x)$ and $\hat{G}(y)$ lie in $(0,1)$. For further discussion of this scaling see \cite{Genest1995} and \cite{Chen2006}. 

The estimation of the model is carried out in two steps: first, by obtaining estimates of the marginals non-parametrically, and second, by estimating the tail-dependence parameters using maximum likelihood. When the marginals are specified parametrically, the estimation is known as the inference functions for the margins (IFM) approach first proposed by \cite{Joe1996}. When the margins are estimated nonparametrically, the method is a two-step semiparametric procedure known as Canonical Maximum Likelihood (CML). For additional details about CML estimation see \cite{Cherubini2004}. Consistency and asymptotic normality of the IFM estimator under a set of regularity condition is shown in \cite{Joe1997}. IFM method is also known to be highly efficient, and we therefore adopt semiparametric IFM as the main estimation method used in this paper.

In order to properly perform IFM and to ensure that our empirical d.f. estimates are based on i.i.d. observations we apply a GARCH(1,1) filter to all return series to remove heteroskedasticity and then use Kolmogorov-Smirnov test to verify uniformity of the resulting margin estimates. As \cite{Ning2009} point out, volatility filtering is also desirable so that to avoid overstating the extent of extreme dependence in the data (see \cite{Poon2004} for additional details).

\section{The recommendations and returns data}
\label{sec::data}

\subsection{IBES recommendations files}

Our recommendations data consist of twelve monthly vintages of Thomson Reuters I/B/E/S summary files issued between January and December 2011, which we obtain from Wharton Research Data Services (WRDS) database. I/B/E/S data are released on the third Thursday of every month, and contain analyst consensus recommendations covering over 4,000 listed and OTC securities. We therefore begin with the full sample of over 45,000 recommendations issued during this period. The focus in this paper is on I/B/E/S \textit{consensus} recommendations, which represent combined opinion of a particular stock by all security analysts tracked by I/B/E/S and include more than 2,700 firms contributing globally. We restrict our attention to consensus recommendations since we aim to assess the predictive ability of security analysts as a group.

For every security in the database, consensus recommendations are obtained by Thomson Reuters as follows: the opinions of individual analysts polled by I/B/E/S are mapped onto a standardized numerical scale, where $1$ - represents a \textit{strong buy} recommendation, $5$ - represents a \textit{strong sell}; $3$ - is a \textit{hold}; $2$ - is a \textit{buy} and $4$ - is a \textit{sell}. A consensus recommendation for a security is the mean of all individual scores.

Since recommendation quality is inversely-related to numerical score, we can expect \textit{smaller} scores to be associated with larger future excess returns under the hypothesis that analyst recommendations have predictive value. To capture upper and lower-tail dependence, the scoring scale needs to be adjusted so that larger scores represent more favourable recommendations. We therefore change recommendation scale so that 1 now represents \textit{strong sell}, 2 is a \textit{sell}, 3 is a \textit{hold}, 4 is a \textit{buy} and 5 is a \textit{strong buy}, with dependence between \textit{larger} scores and returns now indicating stock-picking ability. Note that this change of scale is purely nominal, and has no effect on the recommendation-return relationship in our sample.

\begin{figure}
\label{fig::recdists}
\includegraphics[scale=0.52]{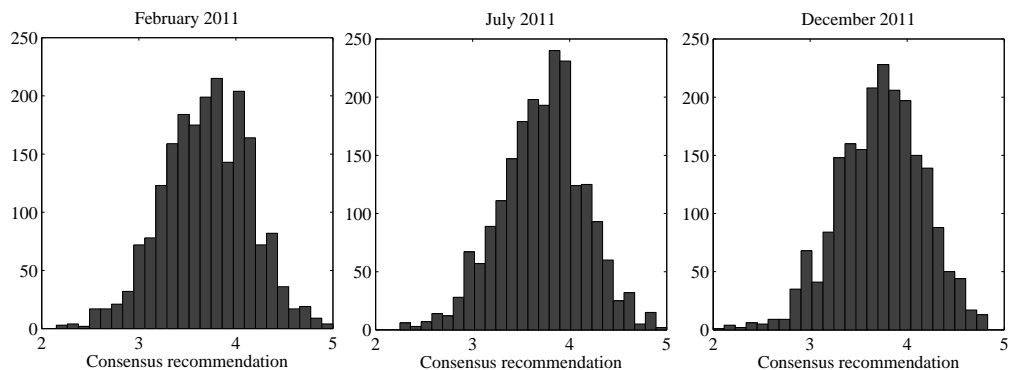}
\caption{Distributions of consensus recommendations in the first, middle \& last I/B/E/S summary files of 2011.}
\end{figure}

The frequency distributions of consensus recommendations at the beginning (January 2011), middle (July 2011) and end (December 2011) of our sample are shown in Figure 1. The shape of the distribution does not change throughout the sample, with skewness toward buy and strong buy recommendations pronounced in all monthly vintages. 

\subsection{Security returns}

For every recommendation in the sample, we obtain subsequent daily closing prices during six months after issue date. We then calculate the corresponding six month excess security returns over a benchmark value-weighted market portfolio, with all price data taken from the S\&P COMPUSTAT database.

Security returns are known to be heteroskedastic, and we first check for the i.i.d.-ness of all daily return series using the ARCH test of \cite{Engle1982}. We find that for most securities in the sample, the null of i.i.d. observations can be rejected. Following \cite{Ning2009}, we filter out heteroskedasticity using a GARCH(1,1) filter, and check again to verify i.i.d.-ness of the filtered series, with securities for which GARCH(1,1) filtering is found insufficient excluded from the sample.

\subsection{Sample restrictions}

Since we are interested in predictive properties of \textit{consensus} opinion, we need to ensure that each mean recommendation indeed represents the view of a sufficiently-large group of analysts. Analyst coverage is extremely skewed toward a small group of large-capitalization firms. In fact, approximately half of all securities in our sample are covered by fewer than 10 analysts \textit{globally} as they usually represent small-capitalization, less-popular or unlisted stocks. For that reason, we exclude all securities recommendations for which are based on fewer than $30$ opinions, and refer to surviving observations as the full sample in the rest of this paper. Additionally, a recommendation that reiterates an earlier opinion may be less informative than a recommendation that represents an outlook upgrade or downgrade. The importance of such recommendation changes is well-documented (e.g. see \cite{Barber2010}). For that reason, we create two sub-samples on which we repeat our estimation: sub-sample A, which contains only those recommendations that represent a downward change in consensus opinion relative to previous I/B/E/S vintage, and sub-sample B, which contains only securities with an upward recommendation change.

\section{The empirical results}
\label{sec::results}

\subsection{Semiparametric estimates of tail dependence}

As a first step, we estimate the marginal distributions of mean recommendations and filtered returns non-parametrically, and perform a Kolmogorov-Smirnov test to ensure that resulting margin estimates are uniformly distributed. We find that we cannot reject the null of uniformity in the full sample as well as in sub-samples, with $p$-values in all cases being very close to $1$. Next, we substitute the empirical d.f. into the SJC copula model and use maximum likelihood to obtain estimates of tail dependence parameters $\lambda_u$ and $\lambda_l$. We first carry out estimation using the full sample (no conditioning on recommendation change), and then repeat using sub-sample A (conditional on recommendation downgrade) and sub-sample B (conditional on recommendation upgrade). 

We find strong evidence of upper, but not lower-tail dependence between consensus recommendation scores and six month excess returns in sub-sample A, but not in the full sample or sub-sample B. Estimation results of the SJC copula model are provided in Table 1.

\begin{table}[htbp]
\label{tbl::empiricalresults}
\begin{center}
\caption{SJC copula estimates: mean recommendation and 6m excess return.}
\begin{tabular}{lccc}
\hline
 & $\hat{\lambda}_u$ & $\hat{\lambda}_l$ & $AIC$ \\ 
 \hline
Unconditional & $0.039$ & $0.006$ & $-14.6$ \\ 
 & $(1.115)$ & $(0.337)$ &  \\ 
Conditional on recommendation upgrade & $0.003$ & $0.001$ & $2.27$ \\ 
 & $(0.420)$ & $(0.001)$ &  \\ 
Conditional on recommendation downgrade & $0.158^{***}$ & $0.014$ & $-19.9$ \\ 
 & $(2.377)$ & $(0.320)$ &  \\ \hline
\multicolumn{ 4}{l}{\small \textit{Notes:} Numbers in parenthesis are $t$-ratios. AIC - Akaike's information criterion. } \\ 
\end{tabular}
\end{center}
\end{table}

The high statistical significance of the estimate of upper-tail dependence coefficient in sub-sample A suggests that the most favorable recommendations tend to be followed by unusually large excess returns, but only when such recommendations represent a decline in consensus opinion compared to previous I/B/E/S file. This may be due to the tendency of investors to over-react in the shorter term to the deterioration of consensus opinion. The lack of significance of $\lambda_l$ indicates that this relationship does not hold for the most unfavourable recommendations, which we find are not associated with greater likelihood of an unusually low excess return in the future. In other words, we find that analysts are able to identify the stocks that will significantly outperform, but not stocks that will underperform, and that their predictive ability depends on the preceding recommendation change.

\section{Discussion}
\label{sec::discussion}
The asymmetries in the relationship between recommendations and returns in the upper and lower tails of the distribution may be due to the difference with which analysts respond to the possibility of under-rating or over-rating a security. Presumably, analyst's objective is to issue recommendations that end up being profitable to the client. For a long-only type investor, a wrong sell recommendation leads to missed profit opportunities, but not to a direct capital loss. This may be much more tolerable than a wrong buy recommendation which leads to monetary losses. Recognizing this, a  risk-averse analyst may require a higher level of confidence to issue an unusually high recommendation than an unusually low one, since making a mistake following the former carries a greater cost.  This would naturally lead to the asymmetric quality of recommendations in the upper and lower tails of the distribution that we document.

\bibliographystyle{elsarticle-harv}
\bibliography{recommendations}







\end{document}